\documentclass[10pt,a4paper, openleft, onecolumn, article] {paper} 
\usepackage{color}
\usepackage{graphicx}
\usepackage{amssymb}

\title{Sequential Sparsening by Successive Adaptation in Neural Populations}

{\small
\author{Farzad Farkhooi$\,^1$, Eilif M{\"u}ller$\,^2$ and Martin
P. Nawrot$^{\,1}$\thanks{Correspondence to: MP Nawrot,
Neuroinformatik, Institut f{\"u}r Biologie, Freie Universit{\"a}t
Berlin, K{\"o}nigin-Luise Stra{\ss}e 1-3, 14195, Berlin |
martin.nawrot@fu-berlin.de}}
}

\begin{document}

\maketitle
\noindent{\footnotesize $^1$~Neuroinformatics and Theoretical
Neuroscience, Institute of Biology, Freie Universit{\"a}t Berlin, and
Bernstein Center for Computational Neuroscience Berlin,
Germany\\$^2$~Laboratory of Computational Neuroscience, EPFL,
Lausanne, Switzerland}

\section{Introduction} 
Sparse coding of sensory input in neural systems is desirable because
it allows for an efficient representation of the environmental scene
at any point in time with only a small number of vocabulary
elements. Moreover, a sparse code of sensory input
may support the efficient formation and subsequent retrieval of
associative memories at later processing stages in the brain
\cite{Foldiak95_895, Baum88_217}. A series of experimental studies in
a number of different model systems have provided evidence for a
sparse stimulus representation in neural spike responses at different
processing stages (for a comprehensive review see
\cite{Olshausen04_481} and references therein).

However, despite extensive theoretical studies of the possible role
and efficiency of a sparse code in neural information processing and
associative memory, the underlying mechanisms that translate a sensory
data into a sparse representation are yet unknown. Among many
possibilities, the mutual roles of network architecture and
interaction of inhibition and excitation are possible candidate
mechanisms that can affect the sparseness of a representation
\cite{Assisi07_1176, Broome06_482,Ganeshina01_335}.

Here, we explore the role of the neuron-intrinsic mechanism of spike
frequency adaptation (SFA) in producing temporally sparse responses to
sensory stimulation at higher processing stages. Our results suggest
that SFA leads to an increased sparseness of the neural response
across successive processing stages in the sense of a temporal
sparseness where the
response becomes increasingly shortened and sharpened. This is
commonly known as {\it lifetime sparseness} \cite{Willmore01_255}.

\begin{figure}[t]
  \begin{center}
    \includegraphics[scale=0.85]{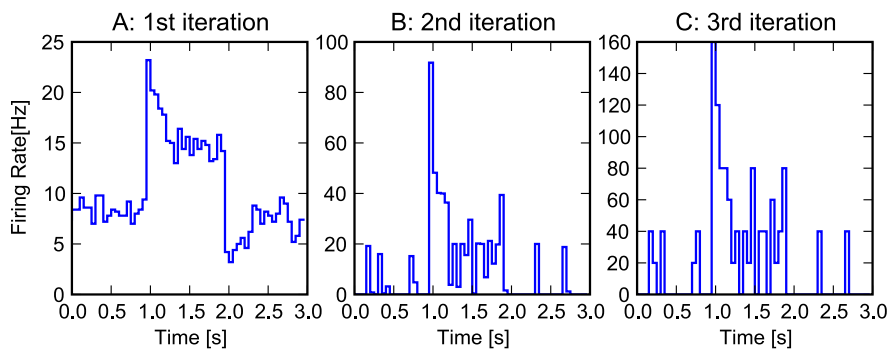}\vspace*{-0.5cm}
  \end{center}
  \caption{\small The effect of iterative spike-frequency adaptation
  in a single cell. A: Spiking activity of a single adaptive
  conductance based neuron \cite{Muller07_2958} averaged across 100
  independent trials, estimated in bins of 50ms size. The neuron is
  driven by Poisson input as described in the text.  B and C:
  Trial-averaged activity of the same neuron model when it is
  stimulated with the output population activity of the previous
  iteration. Simulations performed with pyNEST \cite{Eppler08_12}.}
  \label{fig1}
\end{figure}

\section{Results}

\subsection{Iterative sparsening in a SFA neuron model}

In order to investigate the effect of adaptation on the single neuron
response dynamics, we numerically simulate a five-dimensional master
equation for a conductance-based integrate-and-fire neuron with
spike-frequency adaptation, using the parameters given in
\cite{Muller07_2958}. We then take an iterative approach to mimic a
feed-forward network connection scheme. At the first processing stage,
we provide the model with a step-like Poissonian input
(Fig.~\ref{fig1}\,A spontaneous rate: $1\,$kHz, response rate:
$300\,$Hz, response duration: $1\,$s) and repeat the simulation $100$
times.
%
%
The trial-averaged response is illustrated in the
Peri-Stimulus-Time-Histogram (PSTH) of Fig.~\ref{fig1}\,A. It shows a
clear phasic-tonic response with a strong onset transient and offset
inhibition. We now use the combined output spike trains from all $100$
trials and use this as the single trial input to the same neuron model
at the next processing stage. Again we repeat this complete scenario
for $100$ times and estimate the PSTH in Fig.~\ref{fig1}\,B. We
reiterate once more with $100$ repetitions to estimate the response at
the third processing stage (Fig.~\ref{fig1}\,C).

%
From Fig.~\ref{fig1} we can infer two major results. First, the
response to the initial step-like input turns into a phasic response
that becomes increasingly 'sharp' across repeated iteration. Second,
the response maximum as estimated by the binned firing rate increases
from stage to stage. This response strength naturally scales with the
number of pooled repetitions used for input into the next stage
(simulations not shown). A third but less prominent effect is
expressed in a suppression of spiking during a short period following
the stimulus epoch.




\subsection{Response sparsening in a feed-forward network model}
It has been shown that the primary neurons in the insect mushroom body
- the Kenyon cells (KCs) - use a sparse representation of olfactory
stimuli \cite{Assisi07_1176,Szyszka05_3303} where at each moment in
time only a small fraction of neurons represent the sensory input
(spatial sparseness) and each neuron responds with only a short
response even to long-lasting sensory stimuli (lifetime
sparseness). In a simplified view we may consider the insect olfactory
pathway as a feed-forward network where the olfactory receptor
neurons (ORNs) provide input to the first processing layer of
projection neurons (PNs; approx. 950 uniglomerular PNs in the
honeybee) in the antennal lobe. The PNs project to the layer of KCs
(approx. 160.000 in the honeybee), which in turn provide convergent
input to the mushroom body extrinsic neuron layer (ENs; approx. 400 in
the honeybee).

Here, we designed a reduced feed-forward network with three layers,
 arranged in $10\times100\times10$ lattice with full
 connectivity. This structure represents the processing layers in the
 olfactory pathway: PN$\rightarrow$KC$\rightarrow$EN. All neurons are
 modeled identically using the 5D model and identical parameters as
 described in the previous section. As a means of control we repeated
 the same network simulation but with zero adaptation. To match the
 spontaneous output rates of both models we increased synaptic weights
 by a factor of $1.3$ in the latter.



\begin{figure}[t]
  \begin{center}
    \includegraphics[scale=0.85]{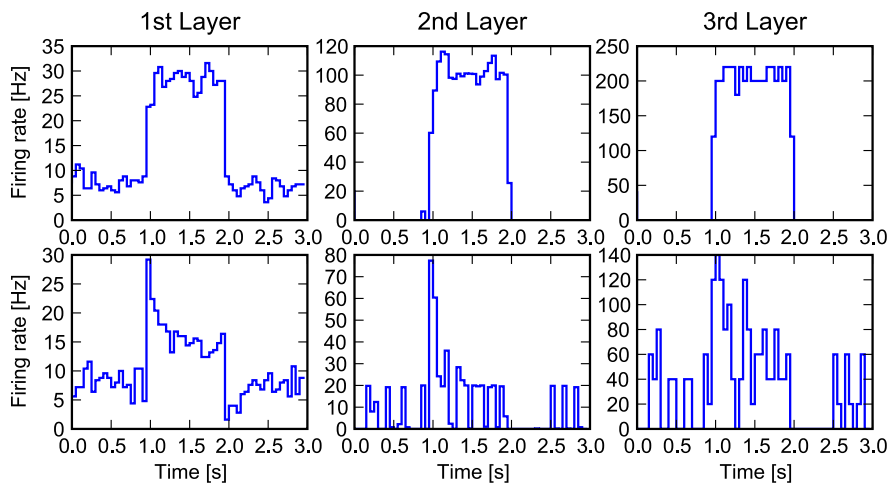}\vspace*{-0.5cm}
  \end{center}
  \caption{\small Simplified network of feed-forward processing in the
  insect olfactory system. Average activity of simulated network for
  10 independent trials in 50ms bins without adaption for control
  (first row) and with SFA (second row), where synapses are assumed to
  produce an $\alpha$ shaped excitatory conductance profile. The
  network simulation was conducted in pyNEST \cite{Eppler08_12}.}
  \label{fig2}
\end{figure}

As ORN input to the PN layer we again used a step-like Poisson input,
independent for all neurons. We then repeated the simulation for 10
independent runs and estimated the population activity for each layers
by constructing the PSTH across all neurons within this layer as shown
in Fig.\ref{fig2}.
%
%
Our result clearly shows that the population response becomes
increasingly phasic as it propagates through the network of SFA
neurons (Fig.~\ref{fig2}, upper row).  However, in the control case
where we switched off SFA in all neurons, the step-like input is
conserved across all layers and the changes in response strength
reflect the neuron number in each layer.


\section{Discussion}


We report here that the simple neuron-intrinsic mechanism of
spike-frequency adaptation can account for the sparse spike response
scheme known as lifetime sparseness \cite{Willmore01_255} in
downstream neurons. SFA reflects high-pass filter properties with
respect to the temporal profile of the neuron's input activity. This
suggests that a feed-forward network with SFA neurons focuses on the
temporal differences of the sensory input. In other words, it suggests
that for higher brain centers it is most relevant to process dynamic
changes in the sensory environment and to neglect the static part of
receptor sensation.

We designed our network model in coarse analogy to the insect
olfactory system. Our model observation match with experimental
observations. At the first processing level of projection neurons the
stimulus response still shows a strong and outlasting tonic part, as
was e.g.\ observed in intracellular recordings from PNs in the
honeybee~\cite{Krofczik09_doi}. The effect fully develops in the
second layer of KCs with a sharp transient response. This matches the
repeated experimental observations in extracellular recordings from
the locust (e.g.~\cite{Broome06_482}) and imaging results in the
honeybee~\cite{Szyszka05_3303}. Finally, the phasic responses in ENs match with experimental
recordings in the honeybee~\cite{Pamir08_doi}.

\section*{Note}
This manuscript was submitted for review to the Eighteenth Annual
Computational Neuroscience Meeting CNS*2009 in Berlin and accepted for
oral presentation at the meeting. A short abstract version was
published as \cite{Farkhooi09_O10}.

\bibliographystyle{abbrv}
\small{
\bibliography{abstarct_long}

\begin{thebibliography}{10}

\bibitem{Assisi07_1176}
C.~Assisi, M.~Stopfer, G.~Laurent, and M.~Bazhenov.
\newblock Adaptive regulation of sparseness by feedforward inhibition.
\newblock {\em Nat Neurosci}, 10(9):1176--1184, 2007 Sep.

\bibitem{Baum88_217}
E.~Baum, J.~Moody, and F.~Wilczek.
\newblock Internal representations for associative memory.
\newblock {\em Biol Cybern}, 59:217--228, 1988.

\bibitem{Broome06_482}
B.~M. Broome, V.~Jayaraman, and G.~Laurent.
\newblock Encoding and decoding of overlapping odor sequences.
\newblock {\em Neuron}, 51(467-482), 2006.

\bibitem{Eppler08_12}
J.~M. Eppler, M.~Helias, E.~Muller, M.~Diesmann, and M.-O. Gewaltig.
\newblock Pynest: A convenient interface to the nest simulator.
\newblock {\em Front Neuroinform}, 2:12, 2008.

\bibitem{Farkhooi09_O10}
F.~Farkhooi, E.~M{\"u}ller, and M.~P. Nawrot.
\newblock Sequential sparsing by successive adapting neural populations.
\newblock {\em BMC Neuroscience}, 10 (Suppl I):O10, 2009.

\bibitem{Foldiak95_895}
P.~Foldiak.
\newblock Sparse coding in the primate cortex.
\newblock In {\em The Handbook of Brain Theory and Neural Networks
  somatosensory cortex}, pages 895--989. MIT Press, 1995.

\bibitem{Ganeshina01_335}
O.~Ganeshina and R.~Menzel.
\newblock Gaba-immunoreactive neurons in the mushroom bodies of the honeybee:
  An electron microscopic study.
\newblock {\em The Journal of Comparative Neurology}, 437(3):335--349, 2001.

\bibitem{Krofczik09_doi}
S.~Krofczik, R.~Menzel, and M.~P. Nawrot.
\newblock Rapid odor processing in the honeybee antennal lobe network.
\newblock {\em Front Comput Neurosci}, 2(9), 2009.

\bibitem{Muller07_2958}
E.~Muller, L.~Buesing, J.~Schemmel, and K.~Meier.
\newblock Spike-frequency adapting neural ensembles: beyond mean adaptation and
  renewal theories.
\newblock {\em Neural Comput}, 19(11):2958--3010, 2007 Nov.

\bibitem{Olshausen04_481}
B.~A. Olshausen and D.~J. Field.
\newblock Sparse coding of sensory inputs.
\newblock {\em Curr Opin Neurobiol}, 14(4):481--487, 2004 Aug.

\bibitem{Pamir08_doi}
E.~Pamir, M.~Schmuker, M.~Strube-Bloss, R.~Menzel, and M.~Nawrot.
\newblock Detection of odor-specific response latencies in the honeybee
  olfactory system.
\newblock {\em Front Comput Neurosci. Conf Abstr: Bernstein Symposium 2008},
  2008.

\bibitem{Szyszka05_3303}
P.~Szyszka, M.~Ditzen, A.~Galkin, C.~G. Galizia, and R.~Menzel.
\newblock Sparsening and temporal sharpening of olfactory representations in
  the honeybee mushroom bodies.
\newblock {\em J Neurophysiol}, 94(5):3303--3313, 2005.

\bibitem{Willmore01_255}
B.~Willmore and D.~J. Tolhurst.
\newblock Characterizing the sparseness of neural codes.
\newblock {\em Network}, 12(3):255--270, 2001 Aug.

\end{thebibliography}
}

\end{document}